\documentclass[12pt,preprint]{aastex}

\usepackage{color}
\definecolor{brown}{rgb}{0.42,0.24,0.07}

\shorttitle{The old open cluster Auner~1}
\shortauthors{Carraro et al.}

\begin{document}


\title{Photometry of a Galactic field at l = 232$^o$, b = -6$^o$.
The old open cluster Auner~1, the Norma-Cygnus spiral arm
and the signature of the warped Galactic Thick Disk}


\author{Giovanni Carraro\altaffilmark{1,2}}
\affil{Departamento de Astronom\'ia, Universidad de Chile,
Casilla 36-D, Santiago, Chile}
\email{gcarraro@das.uchile.cl}

\author{Andr\'e Moitinho\altaffilmark{3}}
\affil{SIM/IDL, Faculdade de Ci\^encias da Universidade de Lisboa, Ed.
  C8, Campo Grande, 1749-016 Lisboa, Portugal}
\email{andre@oal.ul.pt}

\author{Manuela Zoccali}
\affil{Universidad Catolica de Chile, Department of Astronomy \& Astrophysics,
Casilla 306, Santiago 22, Chile}
\email{mzoccali@astro.puc.cl}

\author{Ruben A. V\'azquez}
\affil{ Facultad de Ciencias Astron\'omicas y Geof\'{\i}sicas de la
UNLP, IALP-CONICET, Paseo del Bosque s/n 1900, La Plata, Argentina}
\email{rvazquez@fcaglp.fcaglp.unlp.edu.ar}

\author{Gustavo Baume}
\affil{ Facultad de Ciencias Astron\'omicas y Geof\'{\i}sicas de la
UNLP, IALP-CONICET, Paseo del Bosque s/n 1900, La Plata, Argentina}
\email{gbaume@fcaglp.fcaglp.unlp.edu.ar}

\altaffiltext{1}{Astronomy Department, Yale University,
P.O. Box 208101, New Haven, CT 06520-8101 , USA}
\altaffiltext{2}{ANDES fellow, on leave from Dipartimento di
  Astronomia, Universit\`a di Padova, Vicolo Osservatorio 2, I-35122,
  Padova, Italy}
\altaffiltext{3}{CAAUL, Observat\'orio Astron\'omico de Lisboa, Tapada
  da Ajuda, 1349-018 Lisboa, Portugal}


\begin{abstract}
  We perform a detailed photometric study of the stellar populations
  in a Galactic Field at l = 232$^o$, b = $-6^o$ in the Canis Major
  (CMa) constellation.  We present the first $U,B,V,I$ photometry of
  the old open cluster Auner~1 and determine it to be
  $\approx3.25$~Gyr old and to lie at 8.9~kpc from the Sun.  In the
  background of the cluster, at more than 9~kpc, we detect a young
  population most probably
  associated to the Norma Cygnus spiral arm.  Furthermore,
  we detect the signature of an older population and identify its Turn
  Off and Red Giant Branch. This population is found to have a mean
  age of 7~Gyrs and a mean metallicity of Z$=0.006$.  We reconstruct
  the geometry of the stellar distribution and argue that this older
  population - often associated to the Canis Major {\it galaxy}-
  belongs in fact to the warped old thin/thick disk component along
  this line of sight.
\end{abstract}


\section{Introduction}
In recent papers (Carraro et al. 2005a,  Moitinho et al. 2006, V\'azquez
et al. 2006)  we  have been unveiling   a new detailed picture  of the
structure of   the  Milky Way's  disk in  the   Third  Galactic Quadrant
(3GQ).  Our analysis of  the young  stellar  population and
molecular clouds, which make up the Galactic Thin Disk, has shown that:
({\it i}) the Local Arm, also called the Orion Spur, apparently
enters the 3GQ where it is seen between l=220$^o$ and l=250$^o$.
It seems to remain close to
the formal Galactic plane, b=0$^o$, up to 5 kpc from the Sun where it
starts to descend abruptly, reaching z$=-1.5$ kpc below the plane at
9-11 kpc from the Sun;
({\it ii}) the Local arm does not appear to be a grand design arm ,
but an inter-arm structure, a bridge emerging from the
Carina-Sagittarius arm in the First Quadrant and possibly reaching the
Norma-Cygnus (Outer) arm in the 3GQ;
({\it iii}) the Outer arm in the 3GQ  is  visible from l =
200$^o$ to l = 260$^o$.  We note that this picture bears some
similarity with the one sketched by Moffat et al. (1979), almost 30
years ago;
({\it iv}) The presence of the Local and Outer arms below the b=0$^0$
plane is an effect of the Warp of the Galactic disk (for further
descriptions of the warp see May et al.  1997, Momany et al. 2006,
Moitinho et al. 2006, and references therein);
({\it v}) the Perseus arm (mostly visible in the second quadrant) is
not clearly traced in the 3GQ;
({\it vi}) the young stellar and molecular Warp reaches its southern
maximum at l = 250$^o$-260$^o$.
\noindent
Evidence for the
existence of an old population (4-10 Gyr) in the 3GQ has been reported
by Bellazzini et al. (2004), and has been interpreted as a {\it
  galaxy} - the Canis Major (CMa) {\it galaxy} - undergoing an
in-plane accretion onto the Milky Way. Besides the old population, the
CMa galaxy has been considered to also contain a younger (1-2 Gyr)
component, based on the presence of a blue sequence (popularized as
the {\it Blue Plume} (BP)) in Color-Magnitude Diagrams (CMDs).  But,
as found by Carraro et al. (2005) and confirmed in Moitinho et al.
(2006) and Pandey et al. (2006),
these blue stars are actually very young ($\leq$100 Myr) and
are seen in many different regions of the second and
 3GQ.  Noticeably, most of
them were found to trace the Norma-Cygnus spiral arm in remarkable
agreement with other tracers such as CO molecular clouds and
analytical models of spiral structure.  As for a possible very old
population associated to the CMa {\it galaxy}, which would be similar
to the ones seen in almost all Local Group dwarf galaxies (Mateo
1998), Momany et al. (2006) have clearly shown that the lack of
any extended Blue Horizontal Branch star in its CMD  argue against
its existence in CMa .
Thus, what seems to be left of the CMa {\it galaxy} is a {\it
  relatively metal rich} ($-0.7 \leq [Fe/H] \leq 0.0$) intermediate
age (4-7 Gyr) population (Bellazzini et al.  2004).  However, these
age and metallicity ranges are suspiciously similar to those of the
Galactic old thin or young thick disks (Norris 1999, Bensby et al.
2004).

In this paper, we analyze a field in the constellation of CMa, centered
on the open cluster Auner~1. The observations reveal the signature of
an intermediate age, relatively metal rich population that, as we will
argue, belongs to the warped old thin/young thick disks.

\section{Observational material}

$U,B,V,I$ CCD photometry of the field under study has been obtained at
CTIO, with the 1.0m telescope operated by the SMARTS
consortium\footnote{http://www.astro.yale.edu/smarts/}.  The telescope
hosts a new 4k$\times$4k CCD camera with a pixel scale of
0$^{\prime\prime}$.289/pixel which allows to cover a field of
$20^{\prime} \times 20^{\prime} $ on the sky.  The field was observed
on November 30, 2005, together with the Landolt (1992)  fields
TPhoenix, Rubin~149, PG~0231+006 and SA~95 to calibrate instrumental
magnitudes to the standard system. The night was photometric with an
average seeing of 1.1 arcsec. Data have been reduced using
IRAF\footnote{IRAF is distributed by NOAO, which are operated by AURA
  under cooperative agreement with the NSF.}  packages CCDRED, DAOPHOT
and PHOTCAL following the point spread function (PSF) method (Stetson
1987). An image of the covered area is shown in Fig.~1.  A more
detailed discussion of the data reduction and calibration is presented
in a forthcoming paper (V\'azquez et al. 2006).

\section{The old open cluster Auner 1}

The    star cluster     Auner~1 ($RA  =   07^{h}:04^{m}:16^{s},   DEC=
-19^{o}:45^{\prime}$)   was first detected  by  Auner  et al. (1980)
during a survey of hitherto unknown objects of  various nature. No CCD
study has been  insofar  performed to  our knowledge.  The  cluster is
rather compact and faint, as can  be judged by inspecting Fig.~1.\\

\noindent
We estimated the cluster radius by performing star counts around the
cluster center using our photometric catalog.  We employed the B and I
bands to monitor absorption effects since there seems to be
  some differential reddening toward the cluster. Indeed, the
reddening map of the region from Schlegel et al. (1998) shows the
presence of significant nebulosity on the south-east corner of the
field, and indicates a mean reddening $E(B-V)_{FIRB}=0.60\pm0.15$ for
this line of sight.  From Fig.~2 one can notice how the cluster
clearly stands out up to 2.5 arcmin, then a halo is visible in the I
(but not in the B) band profile up to 3.5 arcmin before reaching the
level of the field. We therefore assign to Auner~1 a radius of
3.0$\pm0.5$ arcmin.
This results to be twice the one listed in the open cluster catalog of
Dias et al. (2001) which was based on visual inspection of the
brightest stars.\\

\noindent
The cluster's fundamental parameters were derived by applying
  the isochrone fitting method using the Padova library of isochrones
  (Girardi et al.  2000), as illustrated in Fig.~3. In the left
panel, we show the CMD of all the stars, whereas in the right panel we
consider the results of the profile analysis and select only those stars
within 3 arcmin from the cluster center.  The cluster clearly emerges
in this panel, whilst completely hidden in the left panel by the
fore/background population.  The Turn Off Point (TO) is located at V
=19.0, (V-I) = 0.9, there is a readily detectable Red Giant Branch
(RGB), and a possible RGB clump of He-burning stars at V = 16.8. With
a difference in magnitude between the TO and the red clump $\Delta V$
of 2.2 mag, Auner~1 would be 3.5
Gyrs old (Carraro \& Chiosi 1994).\\

\noindent
We performed a detailed isochrone fitting analysis, and show here the
best fit, which is achieved for a 3.25 Gyr Z=0.008$\pm0.002$
isochrone, shifted by E(B-V)=0.32$\pm$0.05 (E(V-I) = 0.40$\pm$0.05)
and (V-M$_V$) = 15.75$\pm$0.15.  This places the cluster at 8.9 kpc
from the Sun.  Accordingly, the Galactic Cartesian coordinates in the
right-handed system where the origin is placed in the Sun, the
Galactic center is at (0.0,-8.5,0.0), and X increases toward l =
90$^o$ (Lynga 1982) are X$_G$ = -7.0, Y$_G$ = 5.4 and Z$_G$ =-0.9 kpc,
assuming R$_{GC,\odot}$= 8.5
kpc. The Galactocentric distance is then 15.6 kpc.\\
Interestingly, this cluster falls in an age bin (3 to 4 Gyr, see
Carraro et al. 2005b and Ortolani et al. 2005) where only a few
clusters were known before.  Therefore Auner~1 is a
significant object for our understanding of the age
distribution of the oldest open clusters.

\section{The stellar populations in the field of Auner~1}
The presence of an almost vertical blue sequence in the $V/V-I$ CMD of
Fig. 3 (left panel), known as the Blue Plume (BP), that has already
been detected in other clusters in this Galactic quadrant (Carraro et
al. 2005; Moitinho et al. 2006) and the location of Auner 1, at $\sim
9$ kpc from the Sun and $\sim 1$ kpc below the galactic plane,
stresses the complexity of the stellar populations in this region of
the Galaxy. It is thus mandatory to investigate in more detail the
relation between Auner 1, the BP
population and the various Galactic disk components.\\

\noindent
To provide a quantitative description of what is happening in this
zone we show in Fig.~4 a series of Two Color Diagrams (TCDs)
including all the
stars having $U-B$, $B-V$ and $V$ in the entire observed field.
Each TCD corresponds to a different magnitude bin in teh CMD.  The upper
middle panel is the TCD of the same stars plotted in the CMD. The
upper right panel is a reference diagram where the solid line
represents the intrinsic locus of un-reddened dwarf stars
(Schmidt-Kaler, 1982); the position of some spectral types and their
absolute magnitudes are also indicated. The arrow indicates the way a
star moves in this diagram when some reddening $E(B-V)$ takes place.
The visual absorption is given by
the standard relation $A_V = 3.1 \times E(B-V)$, which was already
found to hold in this region of the Galaxy (Moitinho 2001).  The
dashed line represents the locus occupied by giant stars of similar
spectral types. In the analysis, special attention must be
paid to the overlap region of dwarf, giant and sub-giant stars at
$0.6 < B-V < 1.1$. \\
It is worth mentioning that all the stars plotted in Fig. 4 have
photometric errors $< 0.10$ mag. in all the filters,
a restrictive condition applied to minimize
distortions in the diagrams. \\
The solid tilted line in the TCDs indicates the reddening path.\\

\noindent
We remind that the procedure we are applying has been adapted from one
developed long ago (see Fenkart et al. 1987) and is based on
estimating the average reddening and distance of selected groups of
stars according to the mean absolute magnitude of each group. An
evident advantage of studying the stellar populations in magnitude
bins is the simplicity of the morphology of the respective TCDs in
contrast with the complex appearance of the global TCD (upper middle
panel).

\noindent

\subsection{Results of the method}
The results of the whole procedure are summarized in Table~1.
Spectral types are assigned to the stars according to their
  position in the TCD. Mean distances and the number of stars in each
  spectral range are also reported.  In this Table, $N$ is the number
  of stars and $d$ is the heliocentric distance in kpc.  Fig.~5
  provides a graphical representation of Table~1 to facilitate its
  interpretation. This Figure shows the trend of star counts with
  heliocentric distance for each spectral type range considered in
  Table~1. Star count error bars have been plotted assuming a poisson
  noise distribution. Star counts are not volume-normalized, being the
  sole aim of the Figure to help the reader understand the
  various entries in
  Table~1, and the occurence of different spectral type concentrations
  along the line of sight.

In panel (\textbf{a}), we show the behaviour of dwarf stars of
spectral types from B6 to F0 , in panel (\textbf{b})
the dwarf stars of spectral types from F0 to K0 
together with the probable
K0-K5 Sub Giants (SG) and dwarfs (D). In panel (\textbf{c}) we plot the
same stars as in (\textbf{b}) but considering all the SG and D stars
as dwarfs in deriving their distances and numbers. For this reason
they are indicated with an asterisk.  This exercise is done in order to
understand the effects and consequences
of possible spectral type mis-interpretations.\\
Finally, in panel (\textbf{d}) the giant stars of all spectral
types are shown.\\

\noindent
The synoptic view of  Fig.~5 and Table~1 allows us to derive the
following considerations:\\

\begin{description}
\item $\bullet$ All the stars having spectral type earlier than F0
  mostly identify the thin disk. This
  is an ensemble of early type stars located at different distances along
  the line of sight. According to star counts, the BP stars (earlier
  than A5) are not
  evenly distributed along the line of sight, but they show some
  concentrations at 5, 9.5 and 12 kpc. Beyond 12-13 kpc the
  uncentainties in spectral type and distance derivation do not allow us
  to detect unambigously any structure.  These concentrations are
  compatible with the presence of the Local arm (and 5 kpc
  clump), and the Norma-Cygnus (the 9.5 and 12 kpc clumps) arm in
  this particular Galactic direction.  The occurence of such
  concentrations is typical of spiral arms, whose structure is
  irregular and clumpy;
\item $\bullet$ An interesting point has been raised by the
    referee, whether the BP stars could be in fact Halo A-type
    subDwarfs (sDBA).  There are several reasons that indicate this is
    not the case.  According to Thejll et al. (1997), with a mean
    $M_V$ $\sim$ 4.5-5.0 the sDBA would be located close to the Sun at
    less than 1.5 kpc.  Also, the statistics of
    sDB stars is 0.21 stars $\times deg^{-2}$ (Green et al. 1986),
    much less than the number of BP stars we find in our field.
    Furthermore, one should explain the clear lack of such a
    significant population of stars in the northern Galactic
    hemisphere, since the Halo stars should also be seen there.  The
    same kind of statistical considerations apply to the
    possibility that the BP stars might be Blue Stragglers.  Finally, the
    BP in the field of Auner~1 is similar to the ones analyzed in our
    previous works (Carraro et al.~2005a; Moitinho et al.~2006), which
    we have shown to be excellent spiral tracers. This latter result
    means that the BP must be young, independently of any photometric
    analysis.
\item $\bullet$ The disk dwarfs
   (later than F0, panels ${\bf b)}$ and ${\bf c)}$) show
  increasing concentration from the Sun , and a notorius hole at about
  8-10 kpc. This behaviour is independent of a possible
  mis-classification of dwarf and subgiant stars;
\item $\bullet$ Finally, the thick disk  giants
  (panel ${\bf d)}$) although much less abundant, exhibit the same
  kind of distribution of  disk dwarfs;
\item $\bullet$ The star cluster Auner~1, at a distance of $\sim$ 9
  kpc, lies where the distribution of thick disk stars shows a minimum.
\end{description}

\subsection{The BP, Auner 1 and the Galactic disk components}

The above analysis of the TCDs series makes evident that the vertical
blue sequence identified as the BP population is mostly composed of
late B- and early A-type stars from the group of stars found at
5.0 and $10\pm1.5$ kpc from the Sun.
In brief, the BP stars compose a narrow
band in the CMD, and are young ($ \leq 100$ Myr), with
$0.30<E(B-V)<0.50$ and a distance modulus $15.7<V-M_V<16.3$.  As
the highest density of this type of stars happens between 9 and 11 kpc
from the Sun, we find this result entirely coherent with our previous
ones where we identified the BP stars as tracers of the Norma Cygnus
or outer-arm (Carraro et al. 2005, Moitinho et al. 2006, Baume et al.
2006).  The other nearer component
at 5 kpc, as stated above,
belong most probably to the Local (Orion) arm.
The nature of the BP
stars has been clearly demonstrated in our diagrams because most of
the reddening happens in the first 6.5 kpc from the Sun (see Table 1)
and because the space behind must contain a small amount of dust.
Therefore, the blue stars we see  are simply far
away and for this reason relatively faint.\\

\noindent
Signatures of an old thin or thick disk (sub-giant, giant and
old dwarf stars) populations are also seen in the series of TCDs at
the middle latitude of Auner~1 where we should expect to find a low
density of them. Indeed, the TCDs show that red giant
stars follow a pattern of increasing reddening (moving from
the $13 \leq V \leq 14$ panel to the $18 \leq V \leq 19$ panel)
 ;
for this reason no obvious concentration like a Red Clump structure
does appear. However, when looking towards Auner 1 we are also looking
at the external border of the Galaxy where the number of components of
the thick disk, such as giant and sub-giant stars, start decreasing
quickly. \\
For this reason, the CMDs in Figs. 3 and 4 show at $18 \leq V
\leq 19$ the lower envelope of the giant branch of the thick disk at
$Z\simeq1.0$ kpc below the galactic plane (a reasonable limit for
thick disk components in a warped and flared structure). Below that
magnitude range we continue to see very faint red dwarf stars,
composing the thick disk as well.\\
It is worth emphazising that  this picture does not depend
on the particular photometric error constraints we used. In fact
in Fig.~3 and 6, we relaxed the error constraints plotting all the points
sources showing that the scenario does not change.  \\

\section{Discussion and Conclusions}

In a flat Galaxy, due to the location of the Sun within the crowded
thin disk and to the low spatial density of the thick disk, it is not
expected to detect clear signatures of the thick disk in a
CMD:
i) close to b=$0^o$ there is high contamination from the thin disk.
ii) close to b=$\pm 90^o$ the number of disk (thin and thick) stars is small.
iii) at intermediate longitudes there is less contamination from the
thin disk and an increased number of thick disk stars is present, but the
large spread in distances (and reddening) will not, in general,
produce clear distinctive features in the CMD.

\noindent
However, the situation is different when looking across a warped disk.
Indeed, when facing the warp, the number of thick disk stars is larger
than when looking at b=$\pm 90^o$, and the number of thin disk stars
will be smaller than when looking along the disk (b=$0^o$).  At the
same time, the thick disk stars that follow the warp will be
concentrated at an approximately common distance (when compared to the
Sun-warp distance). These combined effects will produce a recognizable
sequence of thick disk stars in a CMD (which will also include old
thin disk stars of similar age).  Another expected effect is that,
because the thick disk flares, it should still be detected at lower
latitudes where the thin disk is not seen
anymore.\\

\noindent
This is the case of the field we have analyzed in this paper, which
includes the open cluster Auner~1. We have shown that the
cluster is 3.25 Gyr old, and lies at 8.9 kpc from the Sun.\\
In addition to the cluster population, we have detected other
sequences, uniformly extended in our field, indicative of a young and
an old field population.  The occurrence of these young and old
populations in the same field in the 3GQ is not confined to Auner~1.
The F-XMM field discussed by Bellazzini et al. (2004, their Fig.~2)
shows the same features, and also a few open clusters, among which
Tombaugh~1 (Carraro \& Patat),
which is located at the same longitude and only 1  degree to the
south of Auner~1.\\

\noindent
Fig.~3 shows that there is an excess of giant stars, and a hump of blue stars around
V$\sim$18.5 and (V-I) $\sim$0.6, which appears to be the TO of an
evolved population.  To emphasize this point, we show in
  Fig.~6 the CMD of this region which results from removing the open
cluster Auner~1 (all the stars lying 7 arcmin from the cluster center
without any error constraints).  The TO of an older population, with
its RGB and clump, remain and thus are not due to Auner~1.  To
make it clearer, we also show in Fig.~6 a luminosity function (middle
panel), which displays a weak but significant jump at V $\approx$ 18.5
marking the TO of the old
population.\\

Using isochrones to estimate the parameters of this population, we
find that a Z=0.006$\pm$0.003 ([Fe/H]=-0.50) 7$\pm$1.0 Gyrs isochrone (Fig.~7, right panel)
provides a good description of the V $\sim$ 18.5 TO region.
This fit is achieved by shifting the isochrone by E(B-V) =
0.35$\pm$0.10 and (V-M$_V$) = 15.5$\pm$0.5, thus placing the bulk of
this population at about  7.0
kpc from the Sun, consistent with the findings in Table~1 and
Fig.~5.\\
Nevertheless,  we emphasize that this fit is only indicative of the mean
properties of this population. Indeed the detailed analysis presented
in Table~1 clearly shows that the thick disk population starts to be present
at V = 17.0 reaching a maximum at $\sim$ 18.5. This indicates
that the thick disk stars are evenly distributed at all distances
along this particular line of sight.\\

\noindent
Remarkably, this population possesses the typical features (age and
metal content) of the old thin or young thick disk (Norris 1999,
Bensby et al. 2004).\\

\noindent
Further evidence for the Galactic nature of this population is given
in Fig.~6 by the CMDs of fields centered on the open cluster NGC~2414
(upper panel, Moitinho 2001) and Haffner~9 (upper-mid panel, V\'azquez
et al. 2006) at l $\approx$ 232$^o$ (almost the same l as Auner~1 and
Tombaugh~1), but at b = +1$^o$.94 and -0$^o$.6, respectively.  These
fields are well within the putative CMa {\it galaxy} (Bellazzini et
al. 2004, Fig.~1). However they do not show any presence of the young
BP or of the old TO (we call the attention of the reader to how the
blue hump of stars, seen in the lower panels around V$\sim$18.5, is
not present in the upper panels).  This suggests that the old
population is not ubiquitous toward CMa, like the BP is not (Carraro
et al. 2005), but seems to follow the pattern of the warped Galactic
disk.  In fact the CMDs in Fig~6, normalized to the same area, show
that in the case of Tombaugh~1 (lower panel) the BP and old population
are much less abundant than in Auner~1, reflecting the geometry of the
Galactic disk and providing an estimate of the amplitude of the
warp at this longitude.\\
Moreover, a fit to the old background population in the field of
Tombaugh~1 implies it lies at $\approx$ 9.0 kpc from the Sun, a few
kpc more distant than the bulk of the old population in the field of
Auner~1.  Such an increase in distance is expected, due to the warp
which would place increasingly distant stars at increasingly lower
latitudes.

\acknowledgments
The authors
deeply thanks Jorge May for useful discussions.
A.M.  acknowledges support from FCT (Portugal) through
grants SFRH/BPD/19105/2004 and PDCT/CTE-AST/57128/2004.


\clearpage

\begin{deluxetable}{lrccccccccccccccccccc}
\tabletypesize{\scriptsize}
\rotate
\tablewidth{0pt}
\tablecaption{Star counts per magnitude range V, reddening and assigned spectral type in the area of the open
 cluster Auner 1}

\startdata

&& \multicolumn{15}{c}{Dwarf stars of early spectral type}&&&\\
\cline{1-21}
  \noalign{\smallskip}

$\Delta
V$&&\multicolumn{2}{c}{$V<12$}&\multicolumn{2}{c}{$12<V<13$}&\multicolumn{2}{c}{$13<V<14$}&\multicolumn{2}{c}{$14<V<15$}
&\multicolumn{2}{c}{$15<V<16$}&\multicolumn{2}{c}{$16<V<17$}&\multicolumn{2}{c}{$17<V<18$}&\multicolumn{2}{c}{$18<V<19$}&\multicolumn{2}{c}{$19<V<20$}& \\

\noalign{\smallskip} \cline{1-21} \noalign{\smallskip}

E(B-V)&&\multicolumn{2}{c}{0.2}&\multicolumn{2}{c}{0.1}&\multicolumn{2}{c}{0.25}&\multicolumn{2}{c}{0.3} &\multicolumn{2}{c}{0.45}
&\multicolumn{2}{c}{0.50}&\multicolumn{2}{c}{0.6}&\multicolumn{2}{c}{0.65$?$}&\multicolumn{2}{c}{}&\\
 \cline{1-21}
 \noalign{\smallskip}

&      $<Mv>$ & N & d & N & d & N & d & N & d & N & d & N & d & N & d &  N & d & N & d & total \\

\cline{1-21} \noalign{\smallskip}

B6-A0& -0.20&  2 & 1.5 &   &     & 1 & 3.8 & 3 & 5.4 & 6  & 7.2 & 4 & 9.0 & 5 & 14.4 & 5 &  $?$ &  2 & $?$ & 28 \\
A0-A5&  0.8 &  1 & 0.9 & 1 & 1.9 & 2 & 2.4 & 3 & 3.5 & 17 & 4.6 &21 & 5.7 &45 & 9.1  & 45& 12.0& 19 & $?$ & 154 \\
A5-F0&  2.1 &    &     & 5 & 1.0 & 6 & 1.3 &14 & 1.9 & 4  & 2.5 &14 & 3.1 &15 & 5.0  &   &   &    &  & 58 \\

\noalign{\smallskip} \cline{1-21}
 \noalign{\smallskip}
  \cline{1-21}
\noalign{\smallskip}
 \noalign{\smallskip}

&& \multicolumn{15}{c}{Dwarf stars of late spectral type }&&&\\

\noalign{\smallskip} \cline{1-21} \noalign{\smallskip}
 $\Delta
V$&&\multicolumn{2}{c}{$V<12$}&\multicolumn{2}{c}{$12<V<13$}&\multicolumn{2}{c}{$13<V<14$}&\multicolumn{2}{c}{$14<V<15$}
&\multicolumn{2}{c}{$15<V<16$}&\multicolumn{2}{c}{$16<V<17$}&\multicolumn{2}{c}{$17<V<18$}&\multicolumn{2}{c}{$18<V<19$}&\multicolumn{2}{c}{$19<V<20$}&\\
\noalign{\smallskip} \cline{1-21} \noalign{\smallskip}

E(B-V)&&\multicolumn{2}{c}{0.0}&\multicolumn{2}{c}{0.0}&\multicolumn{2}{c}{0.0}&\multicolumn{2}{c}{0.0}
&\multicolumn{2}{c}{0.0}&\multicolumn{2}{c}{0.1}&\multicolumn{2}{c}{0.2}&\multicolumn{2}{c}{0.3}&\multicolumn{2}{c}{0.3}&\\
 \cline{1-21}
 \noalign{\smallskip}
 & $<Mv>$ & N & d & N & d & N & d & N & d & N & d & N & d & N & d &  N & d & N & d & total \\

\cline{1-21} \noalign{\smallskip}

F0-G0&  3.4 & 1 & 0.4 & 5 & 0.6 & 12 & 1.0 &19 & 1.6 & 10 & 2.6 & 36 & 3.6 & 125 & 5.0  & 219 & 6.8  & 216 & 10.4 & 643 \\
G0-K0&  4.9 & 1 & 0.2 & 1 & 0.3 & 11 & 0.5 &31 & 0.8 & 87 & 1.3 & 91 & 1.7 &     &      &     &      & 152 & 5.2 & 374\\
K0-K5&  6.5 &   &     &   &     &  2 & 0.2 &   &     &    &     &    &     &     &      &     &      &     &     & 2\\

\noalign{\smallskip} \cline{1-21}

\noalign{\smallskip}
 \cline{1-21}
 \noalign{\smallskip}
  \noalign{\smallskip}

&& \multicolumn{15}{c}{Dwarf star of late spectral type or sub-giant star candidates}&&&\\

\noalign{\smallskip} \cline{1-21} \noalign{\smallskip}
 $\Delta
V$&&\multicolumn{2}{c}{$V<12$}&\multicolumn{2}{c}{$12<V<13$}&\multicolumn{2}{c}{$13<V<14$}&\multicolumn{2}{c}{$14<V<15$}
&\multicolumn{2}{c}{$15<V<16$}&\multicolumn{2}{c}{$16<V<17$}&\multicolumn{2}{c}{$17<V<18$}&\multicolumn{2}{c}{$18<V<19$}&\multicolumn{2}{c}{$19<V<20$}&\\
\noalign{\smallskip} \cline{1-21} \noalign{\smallskip}

E(B-V)&&\multicolumn{2}{c}{0.0}&\multicolumn{2}{c}{0.0}&\multicolumn{2}{c}{0.0}&\multicolumn{2}{c}{0.0}
&\multicolumn{2}{c}{0.0}&\multicolumn{2}{c}{0.1}&\multicolumn{2}{c}{0.2}&\multicolumn{2}{c}{0.3}&\multicolumn{2}{c}{}&\\
 \cline{1-21}
 \noalign{\smallskip}
 & $<Mv>$ & N & d & N & d & N & d & N & d & N & d & N & d & N & d &  N & d & N & d & total \\

\cline{1-21} \noalign{\smallskip}

G0-K0&  3.9 &   &     &   &     &    &     &   &     &    &     &    &     & 181 & 4.0  &278  & 5.2  &     &     & 459\\
K0-K5&  1.7 &   &     &   &     &  5 & 2.3 & 7 & 3.6 & 42 &  5.7& 48 & 7.9 &  89 & 11.0 & 91  &14.0  &     &     & 282\\

\noalign{\smallskip} \cline{1-21}

\noalign{\smallskip}
 \cline{1-21}
 \noalign{\smallskip}
  \noalign{\smallskip}

&& \multicolumn{15}{c}{Giant star}&&&\\
\cline{1-21}
  \noalign{\smallskip}
   \noalign{\smallskip}
$\Delta
V$&&\multicolumn{2}{c}{$V<12$}&\multicolumn{2}{c}{$12<V<13$}&\multicolumn{2}{c}{$13<V<14$}&\multicolumn{2}{c}{$14<V<15$}
&\multicolumn{2}{c}{$15<V<16$}&\multicolumn{2}{c}{$16<V<17$}&\multicolumn{2}{c}{$17<V<18$}&\multicolumn{2}{c}{$18<V<19$}&\multicolumn{2}{c}{$19<V<20$}&\\
\noalign{\smallskip} \cline{1-21} \noalign{\smallskip}

E(B-V)&&\multicolumn{2}{c}{0.2}&\multicolumn{2}{c}{0.1}&\multicolumn{2}{c}{0.15}&\multicolumn{2}{c}{0.2}
&\multicolumn{2}{c}{0.25}&\multicolumn{2}{c}{0.25}&\multicolumn{2}{c}{0.3}&\multicolumn{2}{c}{}&\multicolumn{2}{c}{}&\\
 \cline{1-21}
 \noalign{\smallskip}

& $<Mv>$ & N & d & N & d & N & d & N & d & N & d & N & d & N & d &  N & d & N & d & total \\

\cline{1-21} \noalign{\smallskip}

G0-K0&  0.8 & 2 & 0.2 &    &     &    &     &     &     &    &      &    &     &    &     &  &   &  &  & 2 \\
K0-K5&  0.3 & 8 & 1.0 & 4  & 2.3 & 8  & 3.5 & 13  & 5.2 & 28 & 7.5  & 35 &12.0 & 29 & 17.0&  &   &  &  & 125 \\
$>$K5&  -0.5  &   &   & 1  & 3.4 & 2  & 5.0 & 2   & 7.5 & 1  & 10.9 &    &     &    &     &  &   &  &  & 6 \\

\noalign{\smallskip}
\cline{1-21}
\noalign{\smallskip}

\enddata
\end{deluxetable}


\clearpage

\begin{figure}
\rotate
\epsscale{1.1}
\plotone{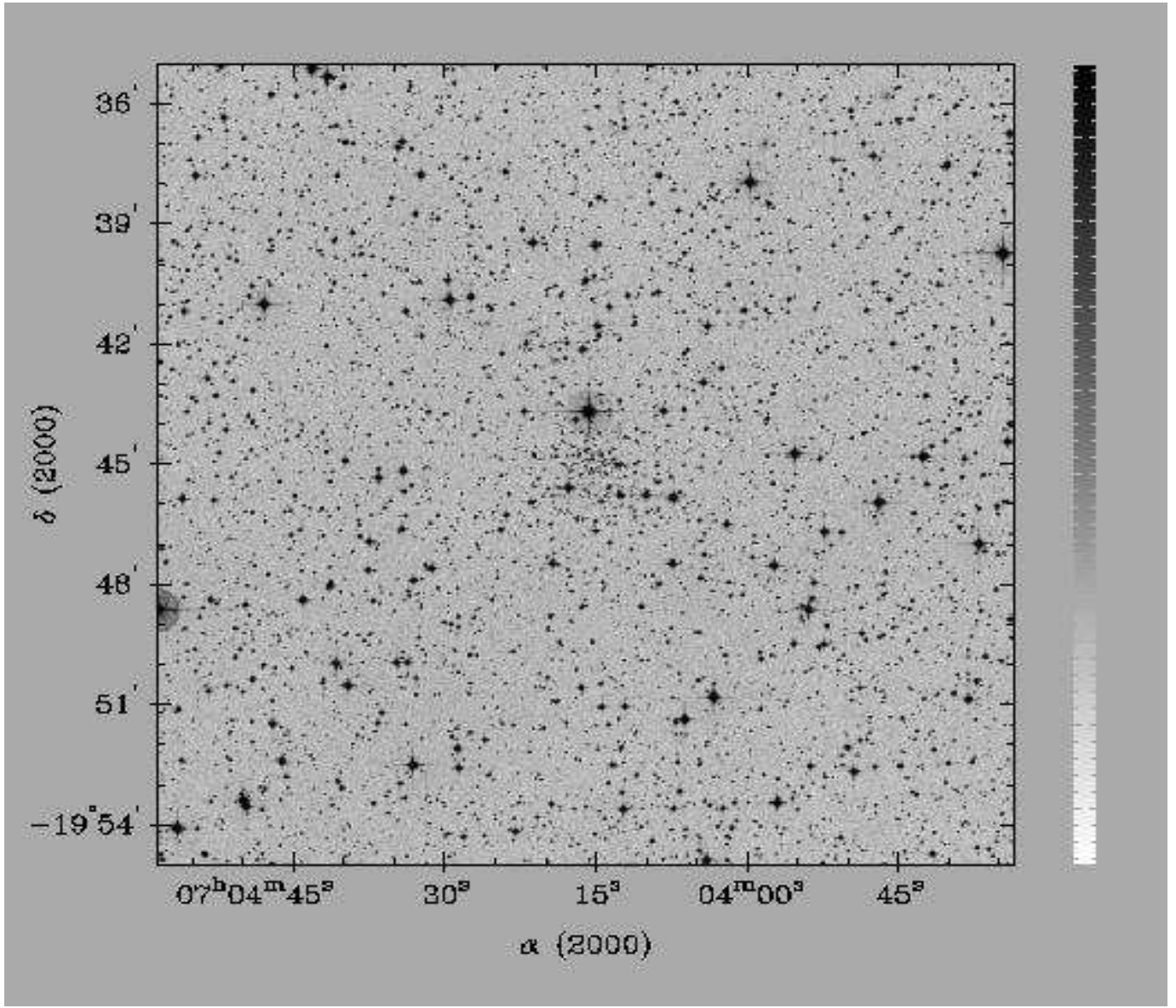}
\caption{A blue Digital Sky Survey image of the area (20 arcmin on a side) discussed in this work.
North is up, East to the left.}
\end{figure}

\clearpage
\begin{figure}
\epsscale{1.0}
\plotone{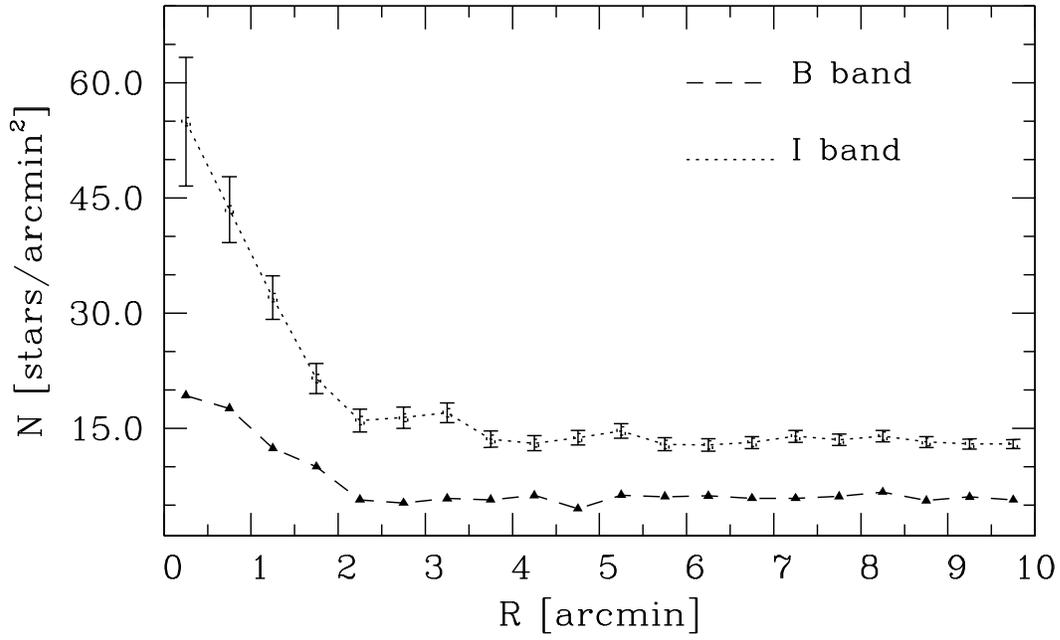}
\caption{Radial density profile of the stars in the field of Auner~1. Star counts
are made by using the B and I band magnitudes, to emphasize absorption effects.}
\end{figure}

\clearpage
\begin{figure}
\epsscale{1.0}
\plotone{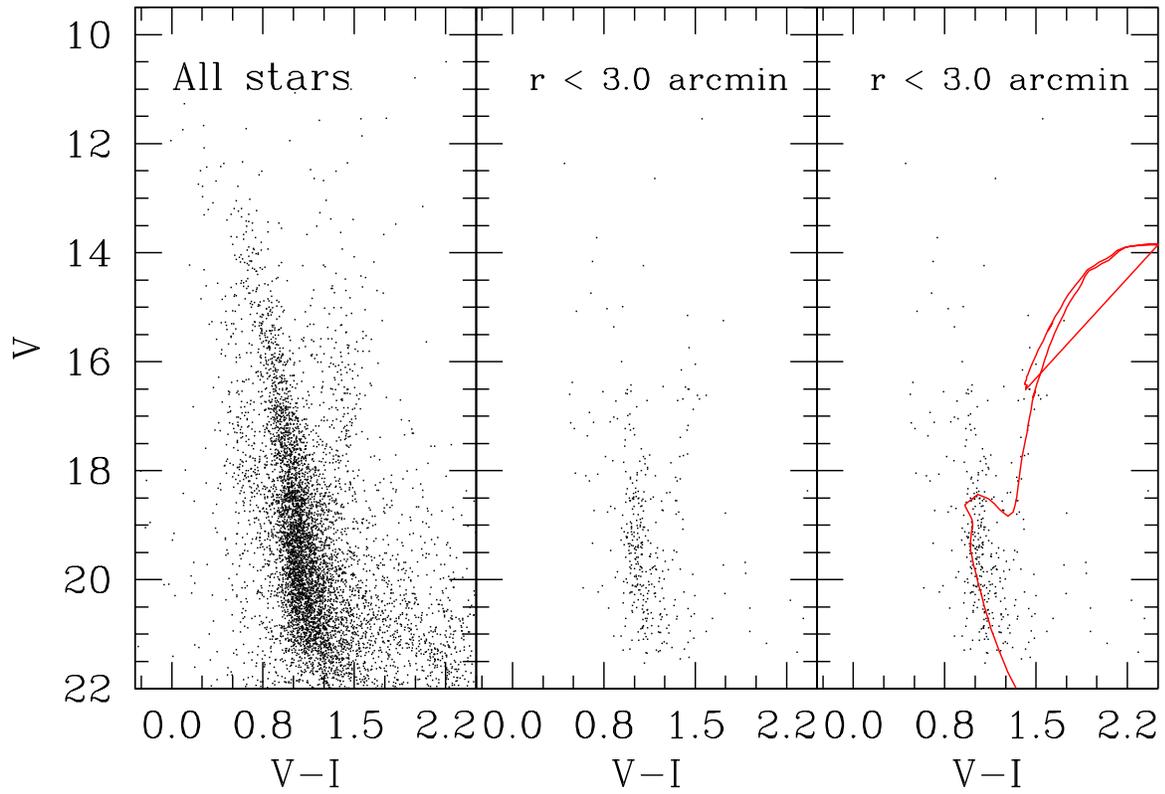}
\caption{CMD of the whole field (left panel), the open cluster Auner~1 (middle panel)
and the iscochrne fitting (right panel). The Z=0.008 isochrone has been shifted by E(V-I)=
0.40 and (V-M)$_V$=15.75. All the stars are plotted without any
constraint on the photometric errors.}
\end{figure}

\clearpage
\begin{figure}
\epsscale{1.0}
\plotone{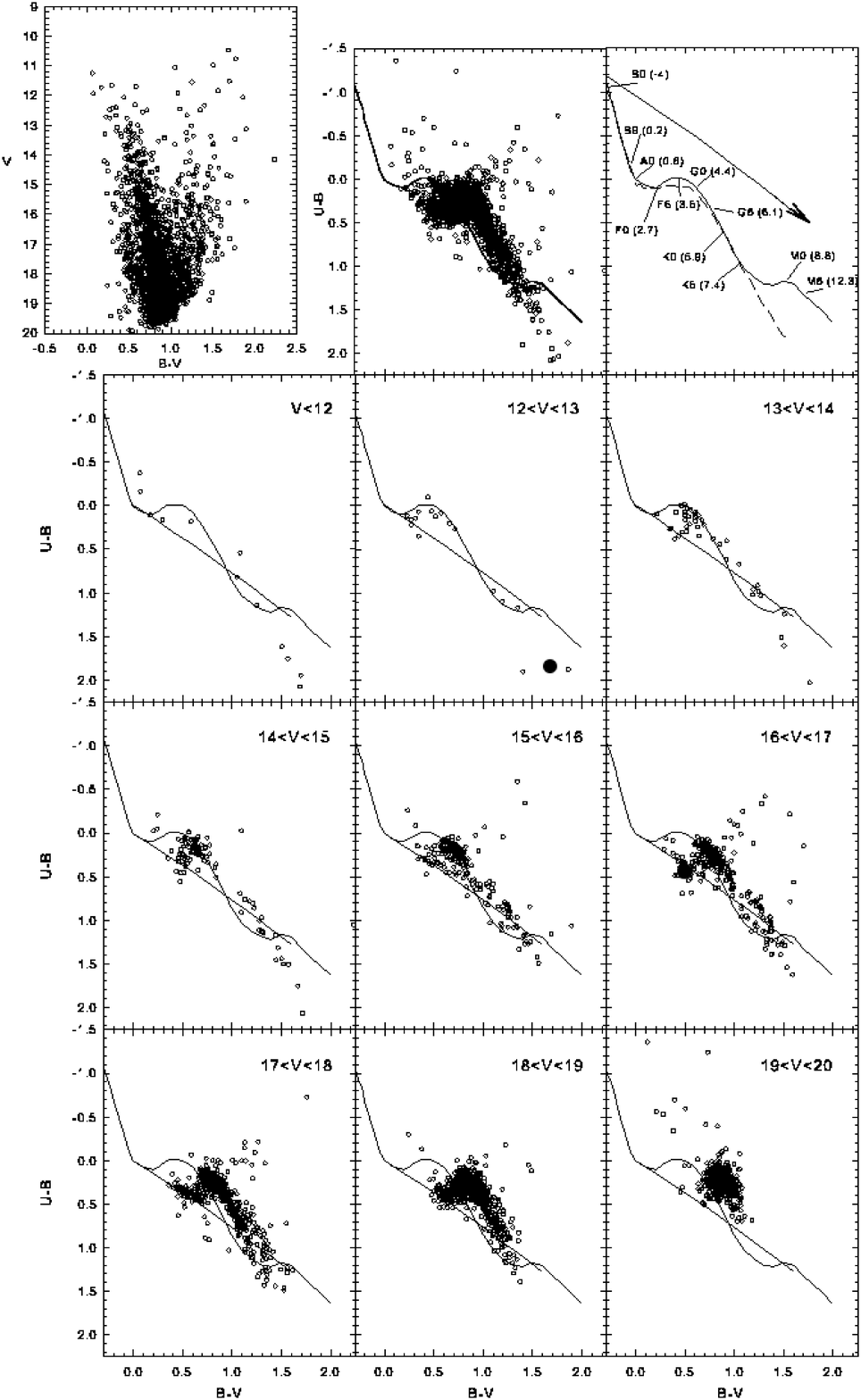}
\caption{Detailed analysis of the population in the field of the open cluster Auner~1. Together
with the cluster CMD, TCDs are shown at varying the V mag. Only stars
having photometric errors lower than 0.1 in all the filters are shown.
See text for details.}
\end{figure}

\clearpage
\begin{figure}
\epsscale{1.0}
\plotone{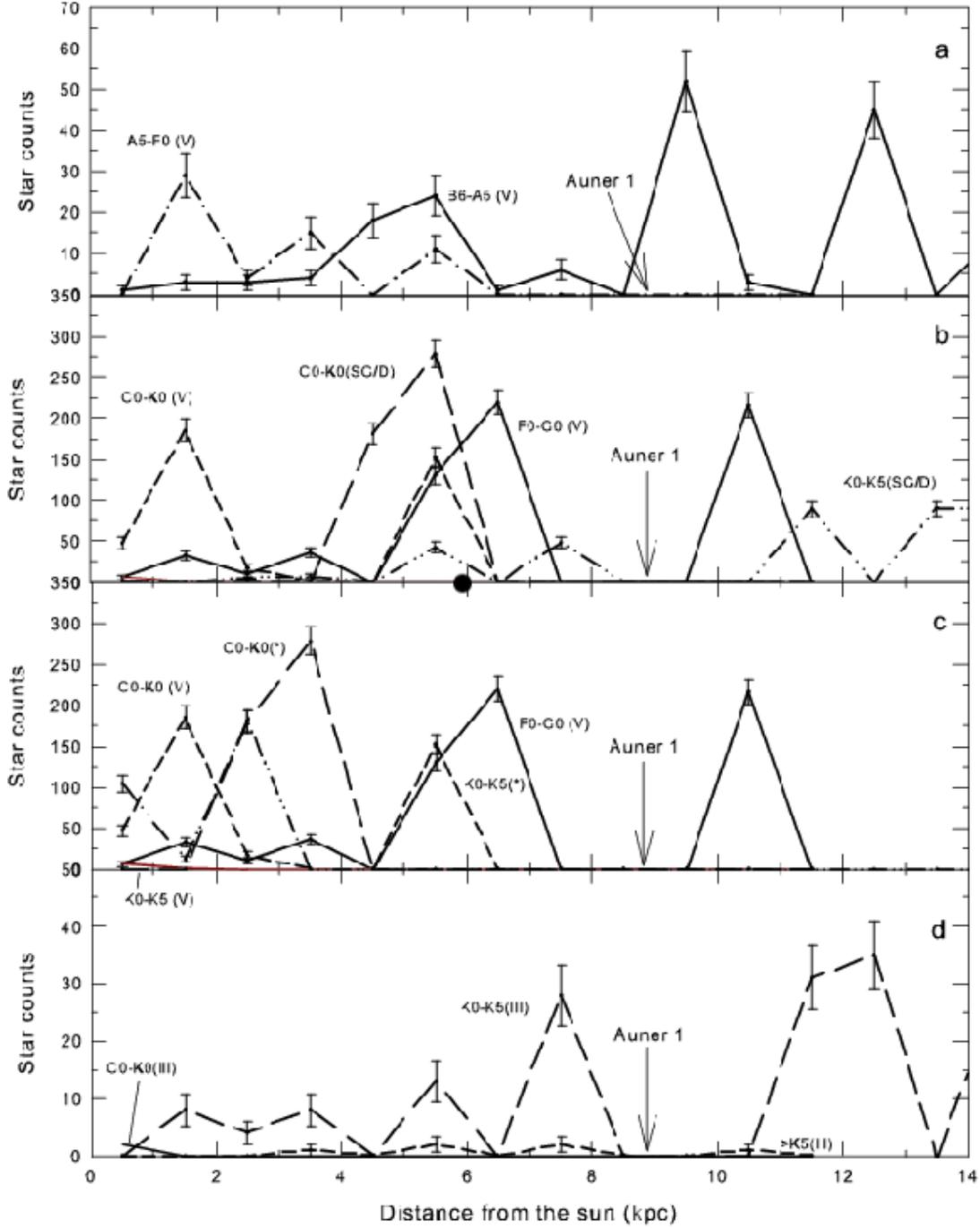}
\caption{Trend of stars counts as a function of heliocentric distance
for stars of different spectral type in the direction of the old
open cluster Auner~1.}
\end{figure}

\clearpage
\begin{figure}
\epsscale{1.0}
\plotone{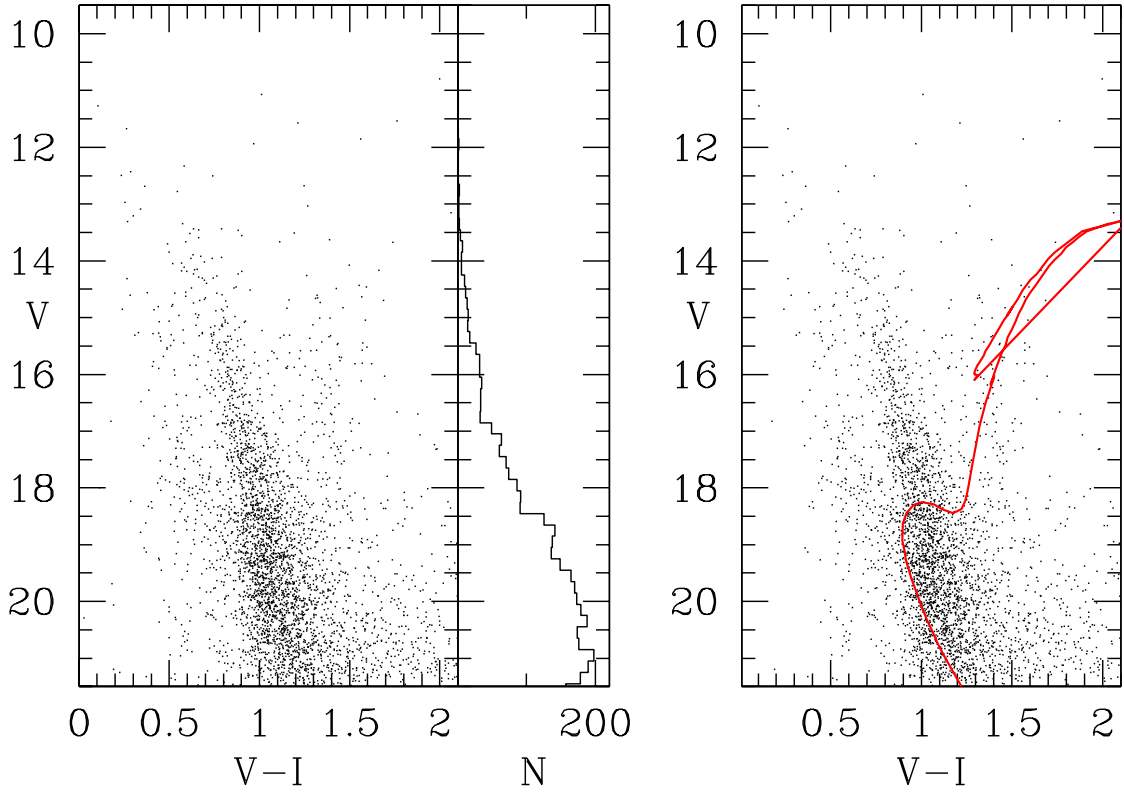}
\caption{The old stellar population in the field of Auner~1. The left panel
shown the CMD of all the stars outside Auner~1 area, the middle panel shows
a luminosity function of the MS, while the right panel presents a possible isochrone
fitting to the old population. The Z=0.006 7 Gyr isochrone has been s
shifted by E(V-I)=0.35 and (V-M$_V$) = 15.5. All the stars are plotted without any
constraint on the photometric errors}
\end{figure}

\clearpage
\begin{figure}
\epsscale{1.0}
\plotone{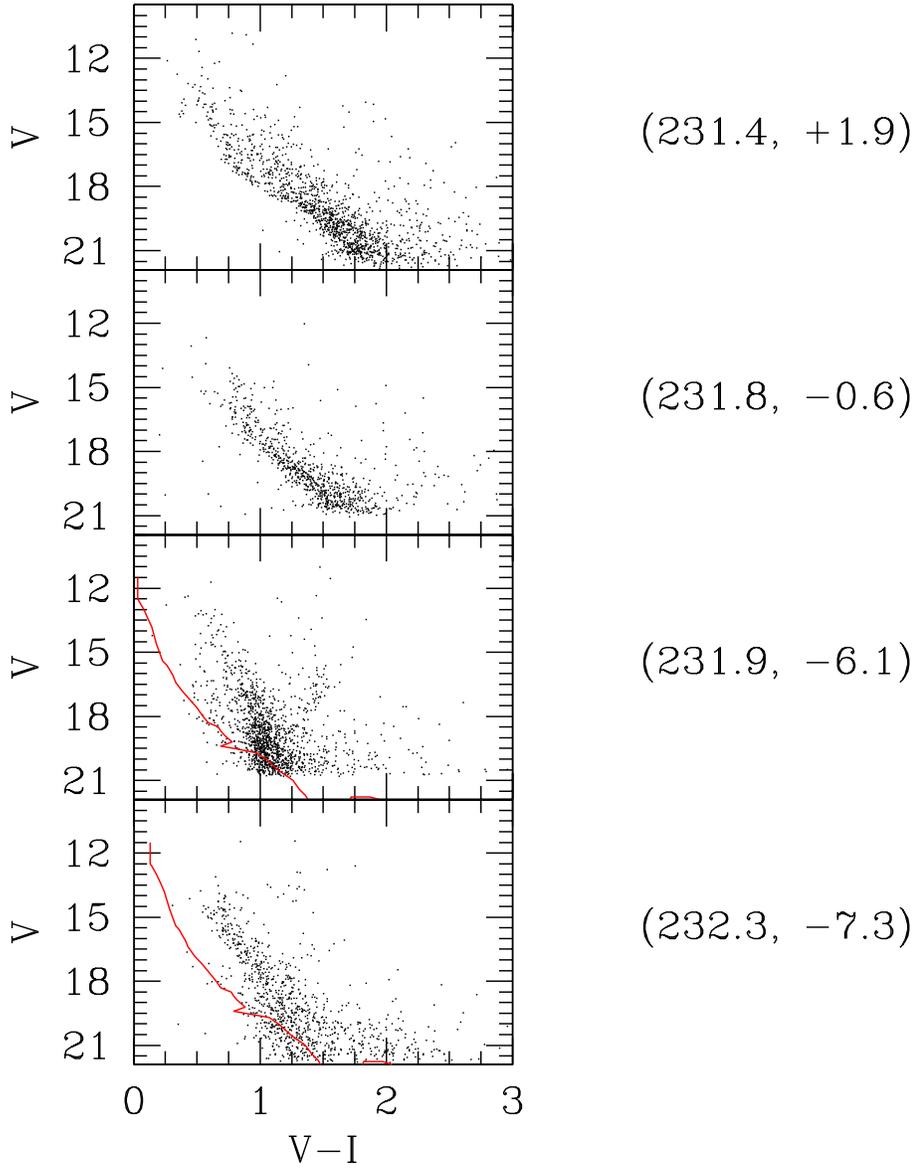}
\caption{The stellar population in a vertical strip at l $\approx 232$.
From the bottom to the top, the area-normalized CMDs of Tombaugh~1, Auner~1,
Haffner~9 and NGC ~2414 are shown. The solid line indicates the locus
of the BP stars}
\end{figure}


\begin{thebibliography}{}
\bibitem[Auner et al.\/(1980)]{aun80} Auner G., Dengel J., Hartl H., Weinberger R., 1980, \pasp~ 92, 422
\bibitem[Baume et al.\/(2006)]{bau06} Baume G., Moitinho A., V\'azquez
        R.A., Solivella G., Carraro G., Villanova S., 2006, \mnras~
        367, 1441
\bibitem[Beers et al.\/(1992)]{bee92} Beers T.C., Preston G.W., Shectman S.A., 1992 \aj~ 103, 1987
\bibitem[Bensby et al.\/(2004)]{ben04} Bensby T., Feltzing S., Lundstrom I., 2004, \aap~ 421, 969
\bibitem[Bellazzini et al.\/(2004)]{mar04}Bellazzini M., Ibata R.A.,  Monaco L., Martin N., Irwin M.J.,
        Lewis G.F.,2004, \mnras~ 354, 1278
\bibitem[Carraro \& Chiosi\/(1994)]{car94}Carraro G. , Chiosi C., 1994, \aap~ 287, 761
\bibitem[Carraro \& Patat\/(1995)]{car95} Carraro G., Patat F., 1995
 \mnras~ 276, 563
\bibitem[Carraro et al.\/(2005)]{car05} Carraro G., V\'azquez R.A., Moitinho A.,
Baume G., 2005a, \apj~ 630, L153
\bibitem[Carraro et al.\/(2005)]{car05} Carraro G., Geisler D., Moitinho A.,
  Baume G., Vazquez R.A., 2005b, \aap~ 442, 917
\bibitem[Dias et al.\/]{dia01} Dias W.S., Alessi B.S., Moitinho A., Lepine J.R.D., 2001, A\&A 389, 871
\bibitem[Fenkart et al.\/(1987a)]{fen87a} Fenkart R., Topaktas L., Boydag S., Kandemir G. 1987 \aaps~ 67, 245
\bibitem[Girardi et al.\/(200)]{gir00} Girardi L., Bressan A., Bertelli G., Chiosi C., 2000, \aaps~ 141, 371
\bibitem[Green et al.\/(1986)]{gre86 } Green R.F., Schmidt M., Liebert J.,
  1986, \apjs~ 61, 305
\bibitem[Lynga\/1982]{lyn82} Lynga G., 1982, \aap~ 109, 213
\bibitem[Mateo\/(1998)]{ma98} Mateo M., 1998, ARA\&A~ 36, 435
\bibitem[Moffat at al.\/(1979)]{mof79} Moffat A.F.J., Jackson P.D.,
  Fitzgerald M.P., 1949, ]aaps 38, 197
\bibitem[Moitinho\/(2001)]{mo01} Moitinho A., 2001, \aap~ 370, 436
\bibitem[Moitinho et al.\/(2006)]{mo01} Moitinho A., V\'azquez R.A., Carraro G.,
Baume G., Giorgi E.E., Lyra W.  2006,\mnras~ 368, L77
\bibitem[Momany et al.\/(2004)]{mom04} Momany Y., Zaggia S.R., Gilmore G.,
     Piotto G., Carraro G., Bedin L., De Angeli F., 2006, \aap~ in press
\bibitem[Norris\/(1999)]{nor99} Norris J.E. 1999, Ap\&SS~ 265, 213
\bibitem[Ortolani et al.\/(2005)]{ort05} Ortolani S., Bica E., Barbuy B., Zoccali M. 2005
    \aap~ 429, 607
\bibitem[Pandey et al.\/(2006)]{pan06} Pandey A.K., Sharma S., Ogura
  K., 2006, MNRAS, preprint
\bibitem[Schlegel et al.(1998)\/]{sch92} Schlegel D.J., Finkbeiner D.P.,
Davis M., 1998, \apj~ 500, 525
\bibitem[Schmidt-Kaler\/(1982)]{sch82}
    Schmidt-Kaler, Th. 1982, Landolt-B\"ornstein, Numerical data and Funct
    ional Relationships in Science and Technology, New Series, Group VI, Vol. 2(b),
    K. Schaifers and H.H. Voigt Eds., Springer Verlag, Berlin, p.14
\bibitem[Thejll et al.\/(1997)]{The97} Thejll P., Flynn C., Williamson
  R., Saffer R., 1997, \aap~ 317, 689
\bibitem[Vazquez et al.\/(2006)]{vaz06} V\'azquez R.A., Carraro G., May J.,
Moitinho A., Bronfmann L., Baume G., 2006, \apj~ submitted
\end{thebibliography}
\end{document}